\documentclass[aps,prc,reprint,superscriptaddress,nofootinbib,longbibliography]{revtex4-2}
\usepackage{amsmath,amssymb,bm,graphicx,booktabs}
\usepackage[colorlinks=true,linkcolor=blue,citecolor=blue,urlcolor=blue]{hyperref}
\newcommand{\Hp}{H^{(+)}}
\newcommand{\cff}{\mathcal{H}}
\newcommand{\Qz}{Q_0^2}
\newcommand{\MS}{\overline{\mathrm{MS}}}
\newcommand{\ReH}{\operatorname{Re}\cff}
\newcommand{\ImH}{\operatorname{Im}\cff}
\newcommand{\dd}{\mathrm{d}}
\begin{document}
\title{Neural reconstruction of generalized parton distributions\\with next-to-leading-order QCD}
\author{Dima Watkins}
\affiliation{Department of Physics, University of Virginia, Charlottesville, Virginia, USA}
\author{I. P. Fernando}
\affiliation{Department of Physics, University of Virginia, Charlottesville, Virginia, USA}
\author{Dustin Keller}
\affiliation{Department of Physics, University of Virginia, Charlottesville, Virginia, USA}
\date{Working draft: October 31, 2026}
\begin{abstract}
We present a neural-network method for reconstructing generalized parton distributions (GPDs) from Compton form factors (CFFs). The network represents quark singlet, nonsinglet, and gluon distributions at a common reference scale. Next-to-leading-order QCD evolution and hard-scattering coefficients connect these distributions to CFFs at the measured scales. Discrete symmetries are imposed exactly, and polynomiality is constrained for a finite set of Mellin moments while leaving their allowed coefficients free. In a noiseless closure test with 45 complex-CFF input positions, the reconstructed charge-weighted quark GPD has a relative $L^2$ error of $4.90\%$ over the evaluation domain. Real- and imaginary-CFF prediction errors at 512 kinematic positions excluded from the fit are $0.75\%$ and $2.75\%$. Tests using only real CFFs at 160 experimental coordinates give GPD errors of $0.75\%$ and $3.05\%$ for two network initializations with exact inputs; these increase to $32.7\%$ and $49.2\%$ for one fixed perturbation at the quoted experimental error scale. For experimentally extracted CFFs, we reconstruct GPDs from the mean of 32 smooth CFF replica functions and from two individual replicas. The resulting GPD shapes depend on both the input fluctuations and the number of optimization iterations. These results demonstrate a working NLO reconstruction method and quantify the contrast between exact-input recovery and reconstruction with uncertain inputs.
\end{abstract}
\maketitle

\section{Introduction}
Understanding the multidimensional quark and gluon structure of the nucleon is a 
central goal of contemporary hadronic physics.
Generalized parton distributions (GPDs), introduced more than two decades ago,
provide a unified framework that simultaneously characterizes the longitudinal
momentum, transverse spatial structure, and spin-orbit correlations of quarks 
and gluons inside hadrons~\cite{Mueller:1998fv,Ji:1996nm, ji-1997-gauge-invariant-decomp}. In deeply virtual Compton scattering (DVCS), the quark GPDs enter through quantum chromodynamics
(QCD) factorization theorems that connect the nonperturbative nucleon structure to the experimentally measurable Compton form factors (CFFs)~\cite{Collins:1998be,Ji:1998xh,Belitsky:2005qn}. The real part of the dominant CFF, $\ReH$, is expressed as a 
singular principal-value (PV) linear integral transform with a singular kernel of the charge-conjugation-even (C-even) quark GPD $\Hp(x,\xi,t,Q^2)$, which makes the inversion problem fundamentally ill-posed \cite{Hansen1998,Tikhonov:1977}. In practice, this linear integral
transform smears localized $x$-space information 
across broad regions of $\xi$ and $t$, causing substantial degeneracy when reconstructing the underlying GPD directly from experimental CFF data \cite{Mueller:2014hsa,Diehl:2002he,Diehl:2003ny}.

Traditional GPD extractions rely on functional parametrizations incorporating
double-distribution models, Regge behaviors, positivity constraints, or
assumptions about polynomiality and $Q^2$ evolution~\cite{Kumericki:2016ehc,Guidal:2013rya,Moutarde:2018kwr}. Current state-of-the-art global fits based on parameterized GPD ans\"atze are consistent with a wide range of experimental and lattice information~\cite{Guo:2025gpd}.
While successful, these methods inevitably introduce model dependence, and their functional rigidity can obscure the true uncertainty associated with the inverse mapping. More recently, machine-learning (ML) approaches have been explored to reduce this bias, most notably in global CFF neural-network analyses of DVCS cross sections and asymmetries~\cite{Kumericki_2011,Cuic_2020,Moutarde:ANN2019,CaleroDiaz_2025}. A milestone in the field applying ML to problems of nucleon structure was the extraction of parton distribution functions (PDFs) --- probability distribution functions arising as the forward limit of GPDs and describing longitudinal momentum distribution of partons in a nucleon --- using artificial neural-networks (ANNs) in the NNPDF project~\cite{Ball:2009determination, Ball:2009determinationerratum, Ball:2013partondist, Ball:2015partondist2, Nocera:2014:unbaisedglobal, DelDebbio:2021get}. An important step toward ANN modeling of
GPDs themselves was taken in Ref.~\cite{Dutrieux:2022ANN}; therein, the authors 
constructed an effectively nonparametric GPD model constrained
by polynomiality~\cite{Ji:1998offforward, RADYUSHKIN199981, Polyakov:1999skewed, Pobylitsa:2003solution}, a requirement imposed by Lorentz invariance, and positivity~\cite{Radyushkin:1998doubledist, Pire:1999positivity, Diehl:2001overlap, Diehl:2001overlaperratum, Pobylitsa:2002inequalities, Pobylitsa:2002disentangling, Pobylitsa:2002positivitybounds, Pobylitsa:2003solution, Pobylitsa:2003integral, Pobylitsa:2004vcspositivity}, reflecting a properly normalized state in the Hilbert space, using pseudo-data generated from phenomenological models.
We reconstruct coupled quark and gluon fields using NLO evolution
and hard matching. To place this calculation in context, we distinguish
the information reconstructed in the different machine-learning
approaches.

Neural CFF analyses reconstruct amplitudes as functions of measured
kinematics, rather than their underlying partonic $x$ dependence
\cite{Kumericki_2011,Moutarde:ANN2019,Le2026}. The lattice-informed
CFF analysis of Huang and Cao also operates at this amplitude
level, incorporating generalized-form-factor information through
dispersion relations \cite{Huang2026}.
Such analyses provide the inputs to a subsequent GPD
reconstruction of the kind studied here. Our experimental application
uses the table of local real-CFF values described in
Sec.~\ref{sec:experiment}, associated with Ref.~\cite{Le2026}, and consequently
inherits its extraction assumptions.

Direct neural GPD modeling is also established.
The construction of Dutrieux \textit{et al.}
\cite{Dutrieux:2022ANN} addresses theoretical constraints in a neural
GPD representation. Our finite-spin moment residuals serve a
different purpose: they constrain polynomiality numerically rather
than building it into the representation.
Dall'Olio \textit{et al.} use Lorentz symmetry and partial
DGLAP-region GPD knowledge to reconstruct the remaining support,
up to a $D$-term, with neural and finite-element methods
\cite{DallOlio2024}. Their input is knowledge of the GPD in a
region, whereas ours consists of CFF convolutions. Riberdy \textit{et al.} study
Bayesian reweighting of a phenomenologically constrained neural
GPD model using mock lattice inputs \cite{Riberdy2024}.
That work provides a complementary route for adding information
to otherwise underconstrained functional directions.

Neural methods using other hadronic inputs further broaden this
context. Chu \textit{et al.} reconstruct distributions from
lattice matrix elements and extend their framework to
zero-skewness GPDs \cite{Chu2026}. Their subsequent
nonzero-skewness study parameterizes generalized form factors
entering Mellin moments \cite{ChuMoments2026}.
Puhan, Sharma, and Kumar combine PDF, form-factor, and lattice
information for pion quark and gluon GPDs \cite{Puhan2026}.
These studies differ from the present proton CFF inversion
in the observables supplied, the hadronic channels, or the
functions directly inferred.

The present field representation uses a deterministic FiLM
residual network: $(\xi^2,t)$ modulates an $x$-dependent
representation whose three outputs describe singlet, nonsinglet,
and gluon fields at a reference scale. Parity and endpoint
suppression are imposed algebraically, and NLO evolution
supplies the scale dependence.
Our nominal inverse objective combines CFF residuals with
finite-spin polynomiality residuals that leave all allowed
moment coefficients free. Closure fits from random initial
weights use Adam steps followed by constrained Gauss--Newton steps
within a trust region, with nonlinear feasibility checks
before acceptance (Sec.~\ref{sec:implementation}).
The experimental family is formed by refitting a deterministic
GPD network to distinct CFF input functions from a common
starting network. Its spread measures the response to
input fluctuations.

GUMP1.0 is an important NLO phenomenological benchmark
\cite{Guo:2025gpd}. It uses a conformal-moment parameterization,
not a neural network, and combines DVCS and meson-production
data with PDF, form-factor, and lattice information.
Our approach instead optimizes a neural field at a reference
scale through the CFF map, with a narrower experimental
input set than GUMP's global analysis.
Likewise, the full-kinematic lattice reconstruction of
Dutrieux \textit{et al.} uses Gaussian-process regression
and double distributions rather than a neural network
\cite{Dutrieux2026}, and is relevant here as an example of
flexible GPD inference from a different type of input.

Against this background, our contribution is the implemented
combination of a conditional FiLM reference-scale field,
ordered NLO evolution and matching, free-coefficient
finite-spin moment constraints, constrained Gauss--Newton
optimization, and a controlled sequence
from exact-input closure to noisy and experimental-input
reconstruction.
Analyses of deconvolution
and shadow GPDs make the distinction between accurate
integral reproduction and field determination particularly
important \cite{Bertone2021,Moffat2023}.
We organize the results around two questions. First, what can the
method reconstruct when the input CFFs are known without noise? Second,
what family of reconstructed functions results when the inputs carry the
reported experimental error scale? Synthetic closure tests answer the
first question; fits to perturbed experimental CFFs address the second.
The experimental application uses previously extracted CFFs rather
than a new fit to cross sections.

Section~\ref{sec:formalism} specifies the fields and QCD map.
Sections~\ref{sec:network} and \ref{sec:implementation} describe the
neural representation and numerical workflow.
Section~\ref{sec:closure} presents exact- and noisy-input closure.
Section~\ref{sec:experiment} gives the experimental reconstruction.
We discuss the interpretation and extensions in
Sec.~\ref{sec:discussion}.

\section{CFFs and their relation to GPDs}\label{sec:formalism}
\subsection{Reference-scale fields and conventions}
We work with the unpolarized vector GPD $H$ at
$\Qz=2~\mathrm{GeV}^2$. For positive $x$, define
$U=H^u(x,\xi,t)-H^u(-x,\xi,t)$ and analogously $D$ and $S$.
The three internal fields are
\begin{equation}
 \Sigma=U+D+S,\quad
 T=U-\frac{D+S}{2},\quad
 H_g=H^g.
 \label{eq:basis}
\end{equation}
\begin{align}
 \Hp&=\frac49 U+\frac19(D+S)
       =\frac29(\Sigma+T).\label{eq:charge}
\end{align}
The gluon convention is $H_g(x,0,0)=xg(x)$.
Both quark fields are extended oddly to negative $x$, whereas $H_g$ is
even. The fields are even under $\xi\to-\xi$.
The reconstructed quantity is the charge-weighted combination
$\Hp$; the reduced basis does not distinguish $D$ from $S$.

At leading order, the real vector CFF contains a principal-value
convolution, and its imaginary part probes the crossover region.
At NLO, quark and gluon coefficients and their evolution must be treated
consistently. We write the complex amplitude in operator form,
\begin{equation}
 \cff(\xi,t,Q^2)
 =\mathcal{C}_{\mathrm{NLO}}(\xi,Q^2)
   \mathcal{U}_{\mathrm{NLO}}(Q^2,\Qz;\xi)
   \bm F_0(\,\cdot\,,\xi,t),
 \label{eq:forward}
\end{equation}
where $\bm F_0=(\Sigma,T,H_g)^T$ and the integrations over
$-1\leq x\leq1$ are included in the operators.
The coefficient functions include the flavor charges once.
The internal real-CFF convention is related to the supplied experimental
table by $\ReH_{\mathrm{int}}=-\ReH_{\mathrm{source}}$.
All comparisons to that table are converted back to its source convention.

\subsection{QCD scale dependence and perturbative accuracy}\label{sec:qcd_accuracy}
A leading-twist GPD parameterizes a hadron matrix element of a
renormalized light-cone quark or gluon operator, with the gauge link
understood. Its factorization-scale dependence follows from operator
renormalization \cite{Diehl:2003ny,Belitsky:2005qn}. A model may specify
the distribution at one scale, but values at different scales must
obey the corresponding QCD evolution. The hard virtuality $Q^2$ and
the factorization scale $\mu_F^2$ are distinct: the latter specifies
the renormalized distribution, even when it is chosen equal to $Q^2$.
Hard matching then connects that distribution to the process amplitude.

We therefore learn the nonperturbative field at a common reference
scale and compute its scale dependence. Every trial network is
evaluated through the same evolution and matching operators before
its CFF residuals are formed. This makes the multi-scale data constrain
one initial field and prevents independent adjustments of its scale
dependence from absorbing CFF discrepancies, and enforces the
evolution equation throughout the fit.

We use fixed $n_f=3$, the $\MS$ scheme, and two-loop
running with $\alpha_s(\Qz)=0.357$.
\mbox{For $\mu_F=\mu_R=Q$}, the evolution equation is
\begin{align}
 \frac{\partial\bm F}{\partial\ln Q^2}
 &=\mathcal{K}_{\mathrm{NLO}}(\alpha_s)\otimes\bm F,\nonumber\\
 \mathcal{K}_{\mathrm{NLO}}
 &=\frac{\alpha_s}{2\pi}\mathcal{K}^{(0)}
  +\left(\frac{\alpha_s}{2\pi}\right)^2\mathcal{K}^{(1)}.
 \label{eq:evolution}
\end{align}
The two-loop kernels follow Ref.~\cite{BFM2000}, and the hard
coefficients follow Ref.~\cite{Moutarde2013}.
Evolution is solved as an ordered integration in $\ln Q^2$, so no
commutation of the leading and next-to-leading generators is assumed.
The singlet and gluon mix, while $T$ evolves in the appropriate
nonsinglet sector.

With $a_s=\alpha_s/(2\pi)$, the hard map retains
\begin{equation}
 \mathcal C_{\mathrm{NLO}}
 =\mathcal C^{(0)}+a_s\mathcal C^{(1)}.
 \label{eq:coefficient_order}
\end{equation}
The gluon coefficient first enters at $O(a_s)$ for this vector CFF.
Our NLO accuracy claim refers to the combined leading-twist
factorization formula: the evolution generator is retained through
$a_s^2$, the hard coefficients through $a_s$, and the running coupling
uses the first two beta-function coefficients. The first omitted
terms in the kernel and hard-coefficient expansions are respectively
$a_s^3$ and $a_s^2$. The coefficients contain the factorization-scale
logarithms required for cancellation of the GPD scale dependence
in the amplitude to the retained accuracy
\cite{BFM2000,Moutarde2013}. The common scheme, flavor content, and
scale prescription are part of this consistency.

The ordered solution resums the scale logarithms generated by the
retained evolution kernels. The neural network determines only the
nonperturbative initial condition; this perturbative content is fixed
throughout training. We choose NLO to include the first hard radiative
corrections and the direct gluon contribution together with the
corresponding evolution accuracy. An NNLO treatment would require the
three-loop evolution kernels, two-loop hard coefficients, and
three-loop running. Power-suppressed corrections are not included.

\subsection{A linear numerical physics map}
For fixed coordinates, coupling, and discretization, the combined
operator in Eq.~\eqref{eq:forward} is linear in $\bm F_0$. We therefore
store its action as numerical rows,
\begin{equation}
 y_{ia}(\theta)=
 \sum_{c,k}W_{ia,ck}\,
 F_{0,c}(x_{ik},\xi_i,t_i;\theta),
 \qquad a\in\{\mathrm{Re},\mathrm{Im}\}.
 \label{eq:rows}
\end{equation}
The rows contain evolution, hard matching, quadrature, and charge
normalization. They can be reused at all $t$ values sharing the same
$(\xi,Q^2)$ because the perturbative kernels do not depend on $t$.
The network is differentiated after contraction with the rows.

The integral map has weakly constrained directions, whose practical
importance grows with finite data, a restricted scale lever arm, and
experimental errors. Closure tests of the field and checks of the
CFFs therefore serve different purposes.

\section{Differentiable neural reconstruction}\label{sec:network}
\subsection{Conditioned reference-scale representation}
The network uses feature-wise linear modulation (FiLM)
\cite{Perez2018} to condition an $x$-dependent representation on
$(\xi^2,t)$. It has width 24, two residual blocks, and 3867 trainable
parameters. With the condition vector
\begin{equation}
 \bm c=\left(\frac{\xi^2-0.18}{0.18},
       \frac{t-\bar t}{s_t}\right),
\end{equation}
where $\bar t=-0.363~\mathrm{GeV}^2$ and
$s_t=0.227~\mathrm{GeV}^2$,
the hidden update has the form
\begin{equation}
 \bm h_{\ell+1}=\bm h_\ell+
 \frac12\tanh\!\left[
 (1+\bm\gamma_\ell(\bm c))\odot
 (A_\ell\bm h_\ell+\bm b_\ell)
 +\bm\beta_\ell(\bm c)\right].
 \label{eq:film}
\end{equation}
Each conditioner is a two-layer network with a tanh hidden activation.
The normalization of $t$ is fixed from the kinematic coordinates.

Let $\bm r_\theta(x,\xi^2,t)$ denote the three raw outputs.
For $c=\Sigma,T$, we set
\begin{equation}
 F_{0,c}=(1-x^2)^2
 \frac{r_{\theta,c}(x,\xi^2,t)-r_{\theta,c}(-x,\xi^2,t)}{2},
 \label{eq:quarknet}
\end{equation}
and use the corresponding sum for $H_g$.
These operations impose parity and endpoint suppression exactly.
There is no $Q^2$ network input: Eq.~\eqref{eq:evolution} supplies all
scale dependence. The envelope fixes the behavior near $x=0$ and
$|x|=1$.

\subsection{Polynomiality with free coefficients}
For even spin $N$, define
\begin{align}
 M_{c,N}(\xi,t)&=\int_{-1}^{1}\dd x\,x^{N-1}F_{0,c}(x,\xi,t),
 &&c=\Sigma,T,\nonumber\\
 M_{g,N}(\xi,t)&=\int_{-1}^{1}\dd x\,x^{N-2}H_g(x,\xi,t).
 \label{eq:moments}
\end{align}
Polynomiality requires an allowed even polynomial in $\xi$ of degree
at most $N$ \cite{Diehl:2003ny,Belitsky:2005qn}. Its coefficients are unknown
generalized form factors. We constrain membership in this polynomial
space while leaving every allowed coefficient free.

At fixed $t_j$, let $\bm m_{cN,j}$ contain moment values at the
chosen skewness nodes. If $V_N$ has columns
$1,\xi^2,\ldots,\xi^N$, an orthonormal basis $Q_N^\perp$ of its
orthogonal complement gives residuals
\begin{equation}
 \bm p_{cN,j}=
 \frac{(Q_N^\perp)^T\bm m_{cN,j}}{s_{cN}},
 \qquad
 J_{\mathrm{poly}}=\frac{1}{Z}
       \sum_{c,N,j}\|\bm p_{cN,j}\|^2 .
 \label{eq:polynomiality}
\end{equation}
Here $s_{cN}=16/[(N+1)(N+3)(N+5)]$ for quarks,
$s_{gN}=16/[(N-1)(N+1)(N+3)]$, and
$Z=n_t n_\xi n_{\mathrm{mom}}$, with 12 channel--spin pairs.
This is equivalent to eliminating
the free polynomial coefficients by least squares.
The nominal fit checks $N=2,4,6,8$ on 41 skewness nodes
between 0.02 and 0.98 at 13 fixed $t$ values.
Moments at different $t$ values do not share fitted coefficients, and
the collocation grid need not coincide with the DVCS measurement
domain. Positivity bounds, individual-flavor sum rules, and the
$D$-term are not constrained by this construction.

\subsection{Loss function and the meaning of a reconstruction}
For the real-only experimental inputs, the objective is
\begin{equation}
 J(\theta)=\frac1n\sum_{i=1}^{n}
  \left[\frac{\ReH_i(\theta)-d_i}{\sigma_i}\right]^2
  +\lambda_{\mathrm{poly}}J_{\mathrm{poly}}(\theta).
 \label{eq:loss}
\end{equation}
We set $\lambda_{\mathrm{poly}}=1$.
Complex-input closure includes both real and imaginary components.
The ideal multiscale lattice uses normalized product quadrature
weights in $(\xi,\ln Q^2,u_t)$ with a residual scale of 0.04 in place
of $\sigma_i$. A known full covariance could replace the diagonal
standardization by a whitening operator; the experimental analysis
uses the quoted marginal uncertainties.

A reconstruction is specified by its inputs, representation, numerical
map, objective, initialization, and stopping rule. Writing the
estimator as $\widehat{\bm F}^{(b)}=\mathcal R_b(\bm d)$ makes
explicit that the number $b$ of optimization iterations is part of the
rule; input replicas probe $\mathcal R_b(\bm d^{(r)})$ for that rule.

\section{Implementation and numerical workflow}\label{sec:implementation}
The calculation separates the costly QCD operator from repeated
network evaluations. The reference field is sampled on continuous
piecewise-polynomial representations with parity extensions and
special treatment of crossover and principal-value structure.
The evolution implementation uses ordered numerical integration.
The source equations, local subtraction terms, and precision guards
are documented with the software~\cite{Keller2026Software}.

Principal-value and plus-distribution terms are assembled through
paired cancellations before evaluating the finite integral.
Near crossover points, selected kernel expressions require higher
precision to avoid cancellation error. The implementation is checked
against local kernel expressions, moment identities, and reference
quadratures.

Closure fits from random initial weights use 400 Adam steps at
learning rate 0.003, followed by up to 60 outer iterations of the
residual-norm constrained solver. Candidate checkpoints are ranked by
the objective alone; the known truth and withheld data are never
used. The solver enforces numerical-resolution and moment conditions
when accepting steps.

Let $\bm r$ denote the stacked objective residuals and
$J_r$ their Jacobian. Trial steps minimize the local
Gauss--Newton model $\|\bm r+J_r\bm d\|^2$ within a
trust radius and the current derivative span.
For a normalized moment-feasibility residual $\bm e$, the
step model retains the condition
\begin{equation}
 \|\bm e+J_e\bm d\|_2\leq1 .
 \label{eq:step}
\end{equation}
Keeping the norm of the affine residual retains curvature
that would be lost by replacing it with a single tangent
halfspace. Other active numerical conditions enter through
their local inequalities, and the original nonlinear conditions
are evaluated before a step is accepted.

All calculations stop at prescribed iteration limits. The experimental
replica fits in Sec.~\ref{sec:experiment} initialize from a common GPD
network fitted to a smooth CFF representation of the unperturbed local
data, and each uses 5, 10, or 20 additional outer iterations. The
common starting network removes initialization differences between
replicas.

At the 160 experimental coordinates there are 102 distinct
$(\xi,Q^2)$ pairs, allowing row reuse. Analytic differentiation is
applied to already contracted CFF and moment outputs, reducing the
need to retain large node-by-parameter derivative tensors.
The ideal closure run required approximately 7.4 hours and
7.1~GiB peak memory on a single workstation.

\section{GPD closure with exact and noisy inputs}\label{sec:closure}
\subsection{A truth independent of the inverse ansatz}
The closure tests use an analytic generating GPD that is independent
of the neural parameterization: a three-channel, $t$-dependent
construction with nonzero ERBL and DGLAP support and calculable
polynomial moments. For channel $c$ it has the form
\begin{equation}
 F^{\mathrm{true}}_{0,c}
 =p_c(x)(1-x^2)^2
  \sum_{i,k}a_{cik}e^{b_{cik}t}
  P_i^{(2,\nu_c)}(2x^2-1)\,\xi^{2k},
 \label{eq:truth}
\end{equation}
where $(p_c,\nu_c)=(x,1/2)$ for quarks and $(1,-1/2)$
for the gluon. The allowed generator terms satisfy $k\leq i+1$.
Jacobi orthogonality supplies the required moment structure.
The coefficients are listed in Appendix~\ref{app:truth}; this
generator has no additional $D$-term.

The ideal input lattice has five skewness values, three scales, and
three physical-$t$ chart positions, giving 45 complex CFFs.
Its envelope is
$0.102\lesssim\xi\lesssim0.446$,
$1.51\leq Q^2\leq8.51~\mathrm{GeV}^2$, and
$-1.37\leq t\leq-0.11~\mathrm{GeV}^2$, with the actual upper
$t$ boundary adjusted at each $\xi$:
\begin{align}
 t(\xi,u)&=t_{\min}
     +u\,[t_{\mathrm{hi}}(\xi)-t_{\min}],\nonumber\\
 t_{\mathrm{hi}}(\xi)&=\min\left[t_{\max},
  -\frac{4M^2\xi^2}{1-\xi^2}-0.01~\mathrm{GeV}^2\right].
 \label{eq:chart}
\end{align}
The margin implements a leading-power kinematic chart.
A separate 512-position CFF check uses skewness values, logarithmic
scale midpoints, and $u$ midpoints absent from the fitting lattice.

We measure field recovery by
\begin{equation}
 \epsilon_H(\Omega)=
 \left[
 \frac{\int_\Omega \dd\mu\,|\widehat{\Hp}-\Hp_{\rm true}|^2}
      {\int_\Omega \dd\mu\,|\Hp_{\rm true}|^2}
 \right]^{1/2}.
 \label{eq:error}
\end{equation}
The wider-chart quadrature uses 33 skewness nodes, nine $u$ nodes,
and order-160 $x$ quadrature in each region. The measure
in Eq.~\eqref{eq:error} gives equal weight to each $(\xi,u)$
node and Gauss weights in $x$. ERBL and DGLAP errors are also
evaluated separately.

\subsection{Recovery from noiseless multiscale CFFs}
\begin{figure*}[t]
 \includegraphics[width=\textwidth]{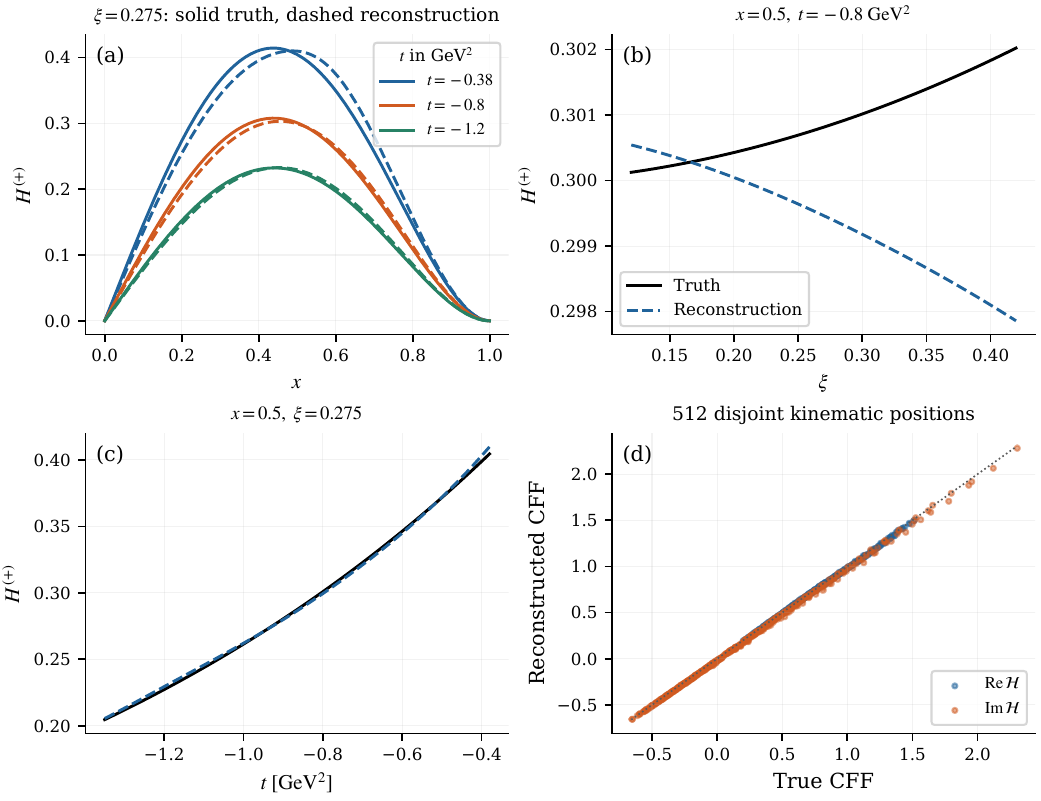}
 \caption{Noiseless multiscale closure at $\Qz=2~\mathrm{GeV}^2$.
 (a) Charge-weighted quark GPD at three momentum transfers and
 $\xi=0.275$; solid curves are the truth and dashed curves the
 reconstruction. (b) Skewness dependence at $x=0.5$ and
 $t=-0.8~\mathrm{GeV}^2$, displayed only up to $\xi=0.42$ to remain
 within the leading-power physical domain. (c) Momentum-transfer
 dependence at $x=0.5$, $\xi=0.275$.
 (d) Complex-CFF predictions at 512 disjoint positions.
 Positive $x$ is shown; the negative-$x$ quark field follows by oddness.}
 \label{fig:exact}
\end{figure*}

\begin{figure*}[t]
 \includegraphics[width=\textwidth]{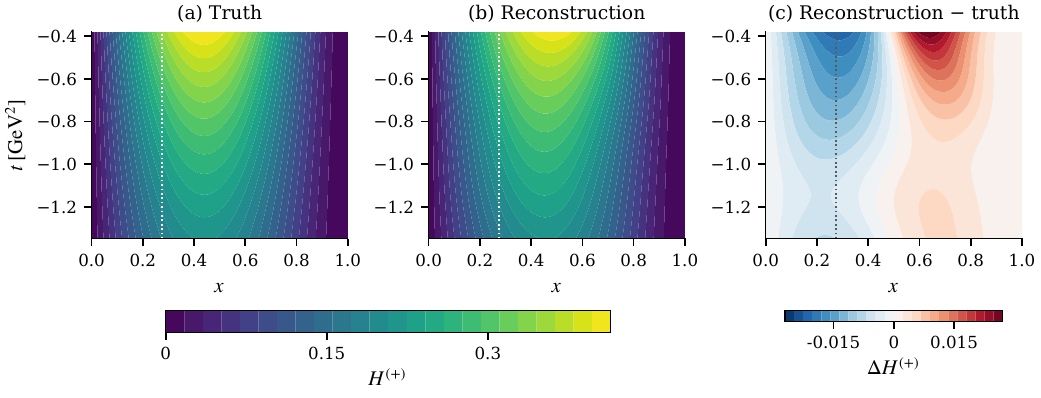}
 \caption{Reference-scale charge-weighted GPD at $\xi=0.275$ in
 the exact-input closure: (a) truth, (b) reconstructed field, and
 (c) their signed difference. The first two panels use a common color
 scale; the third has its own scale. The dotted line marks $x=\xi$.}
 \label{fig:exact_surface}
\end{figure*}

Figure~\ref{fig:exact} shows the reconstructed $x$, $\xi$, and $t$
dependence and the disjoint CFF predictions.
The wider-chart error is $4.90\%$, with $5.41\%$ in ERBL and
$4.79\%$ in DGLAP. The largest sampled slice error is $8.64\%$.
The smooth variation in $t$ and the main $x$ shape are recovered,
while the weaker skewness dependence remains visibly less accurate.
Figure~\ref{fig:exact_surface} displays the field and its signed
difference from the truth over an $(x,t)$ plane.


\begin{table}[t]
 \caption{Noiseless multiscale closure. Relative $L^2$ errors
 are percentages. Disjoint refers to the 512-position
 check outside the fitting lattice.}\label{tab:exact}
 \begin{ruledtabular}
 \begin{tabular}{lr}
 Quantity & Error (\%)\\ \hline
 $\Hp$, wider chart & 4.90\\
 $\Hp$, ERBL / DGLAP & 5.41 / 4.79\\
 $\ReH$, fitted / disjoint & 0.93 / 0.75\\
 $\ImH$, fitted / disjoint & 1.60 / 2.75\\
 $\Sigma$, wider chart & 78\\
 $T$, wider chart & 235\\
 $H_g$, wider chart & 5.90
 \end{tabular}
 \end{ruledtabular}
\end{table}

Table~\ref{tab:exact} also shows that the individual $\Sigma$ and $T$
errors are much larger than the error of their charge-weighted sum,
so the recovery of $\Hp$ does not imply flavor separation. The gluon
error is small in this closure test.

Refining the numerical CFF map changes these predictions at relative
levels of a few parts in $10^6$, well below the fitted CFF errors.

\subsection{Real-only inputs at experimental coordinates}
The next comparison uses the same truth at the actual 160 retained
experimental coordinates. Only $\ReH$ enters the loss.
One input is exact; the other adds a fixed Gaussian perturbation
with the quoted combined experimental error scale.
Both are fitted from two random initializations under the same
protocol, with the noise vector shared between them.

\begin{table}[t]
 \caption{Charge-weighted GPD wider-chart errors at the 160
 experimental coordinates, using real CFFs only. The two seeds
 share the same noisy input vector.}\label{tab:noisy}
 \begin{ruledtabular}
 \begin{tabular}{lrr}
 Initialization & Exact input (\%) & Noisy input (\%)\\ \hline
 1 & 0.75 & 32.7\\
 2 & 3.05 & 49.2
 \end{tabular}
 \end{ruledtabular}
\end{table}
\begin{figure*}[t]
 \includegraphics[width=\textwidth]{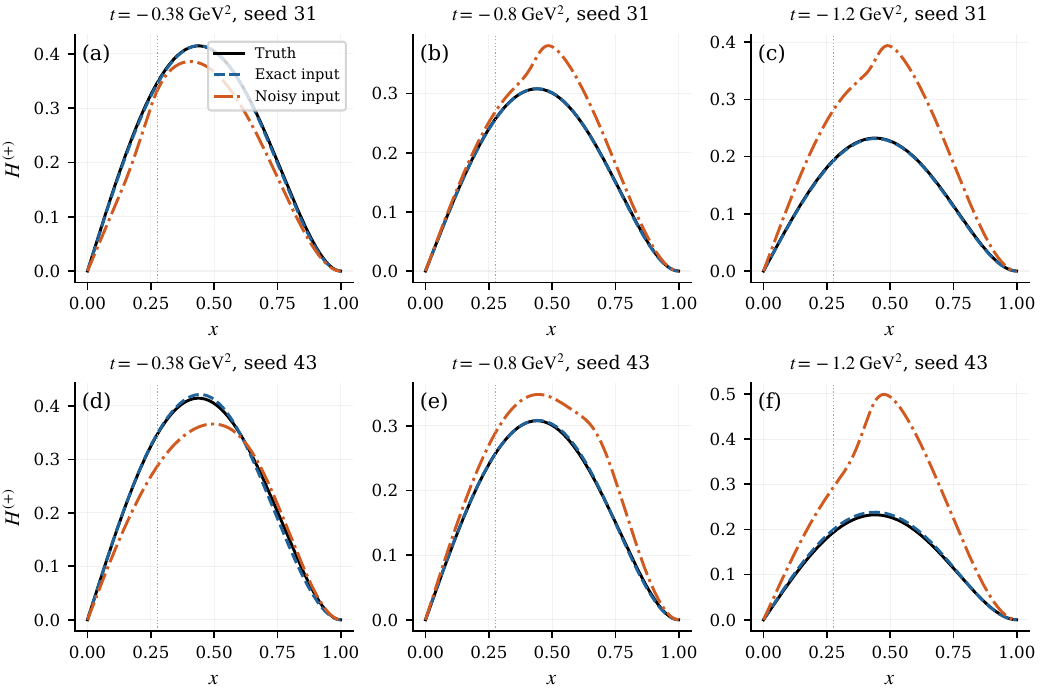}
 \caption{Real-only closure at experimental coordinates.
 The rows correspond to two network initializations; the columns
 show three $t$ values at $\xi=0.275$, $\Qz=2~\mathrm{GeV}^2$.
 Black curves are the same generating truth, blue dashed curves
 use exact CFF inputs, and orange dash-dotted curves use one fixed
 noisy CFF vector. Dotted vertical lines mark $x=\xi$.
 Differences between rows show the initialization sensitivity.}
 \label{fig:noisy}
\end{figure*}

The exact-input errors of $0.75\%$ and $3.05\%$ demonstrate
successful reconstruction for this generating field even with
real-only observations. The noisy counterparts increase to
$32.7\%$ and $49.2\%$.
Figure~\ref{fig:noisy} shows that the change affects the field shape,
not simply an overall normalization.

The ideal-complex and real-only fits have different sampling
geometries, numbers of observations, and objective weights, so the
smaller exact-input error of the latter does not reflect a benefit
from discarding imaginary data.

\section{Reconstruction from experimental CFFs}\label{sec:experiment}
\subsection{Experimental inputs and a fit to local CFF values}
The experimental input is the table of local real-CFF extractions
of Ref.~\cite{Le2026}, using the CDNN or FQDNN value selected
in that analysis. The cuts $Q^2>1.5~\mathrm{GeV}^2$ and $|t|/Q^2\leq0.25$
retain 160 of 189 rows. These span 67 distinct skewness values,
102 $(\xi,Q^2)$ pairs, and 79 momentum transfers.
The combined statistical and systematic uncertainties quoted there
are used as $\sigma_i$. Only $\ReH$ is fitted; $E$, $\widetilde H$,
and a flavor decomposition are not extracted.

The \emph{local-CFF fit} uses these
160 local $\ReH$ values without added perturbations or
intermediate CFF smoothing. This fit uses 13 fixed $t$ nodes
for the polynomiality constraints and has an RMS real-CFF
residual of 2.43 in units of the quoted errors.
The fits described below instead use smooth CFF functions
constructed from perturbed versions of the same table.

\subsection{From perturbed observations to coherent CFF functions}
Uncertainty propagation starts by perturbing the full 160-entry
local CFF vector once per replica:
\begin{equation}
 d_i^{(r)}=d_i+\sigma_i z_i^{(r)},\qquad
 z_i^{(r)}\sim\mathcal N(0,1).
 \label{eq:replicas}
\end{equation}
Each draw is fitted with the same smooth
$3$--$32$--$32$--$1$ tanh CFF network. Its fixed
training/validation split and model seed are retained;
target standardization, validation-epoch selection, and the
prescribed all-row refit are repeated for each draw.
Each CFF replica is therefore one complete function of
$(\xi,t,Q^2)$, preserving correlations induced by the
smoothing operation.

The median pointwise standard deviation of the 32 CFF functions,
divided by the original quoted uncertainty, is 0.48, reflecting the
smoothing. The original fitting weights are retained for the GPD
fits.

We next fit a GPD to the pointwise mean of these 32 CFF
functions, and separate GPDs to the first two CFF replicas,
labeled 1 and 2. All three fits use the same
initial network specified in Sec.~\ref{sec:implementation}.
We compare their outputs after 5, 10, and 20
additional outer optimization iterations from that common
starting network.

The \emph{replica-mean CFF fit} denotes the GPD obtained by
fitting the mean CFF function, as distinct from an average of
reconstructed GPDs.

\subsection{Comparison of reconstructed GPDs}
Table~\ref{tab:experimental_runs} compares CFF reproduction
for the four GPD fits. All residuals use the same 160
coordinates and quoted errors. The replica-mean CFF fit
has an RMS residual of 1.33 against its own CFF target
and 2.17 against the unperturbed local values.
The two individual replica fits give local-data residuals
of 2.19 and 2.13, compared with 2.43 for the
local-CFF fit.

\begin{table}[t]
 \caption{Real-CFF reproduction by the experimental GPD fits.
 Each entry is an RMS residual divided pointwise by the
 original quoted errors. The second column uses each run's
 own CFF target; the third uses the same unperturbed local
 CFF values for all runs. The mean-CFF and replica runs are
 evaluated after 20 additional outer iterations.}
 \label{tab:experimental_runs}
 \begin{ruledtabular}
 \begin{tabular}{lrr}
 GPD fit & Own target & Local CFF values\\ \hline
 Local CFF values & 2.43 & 2.43\\
 Replica-mean CFF & 1.33 & 2.17\\
 CFF replica 1 & 1.53 & 2.19\\
 CFF replica 2 & 1.58 & 2.13
 \end{tabular}
 \end{ruledtabular}
\end{table}

\begin{figure*}[t]
 \includegraphics[width=\textwidth]{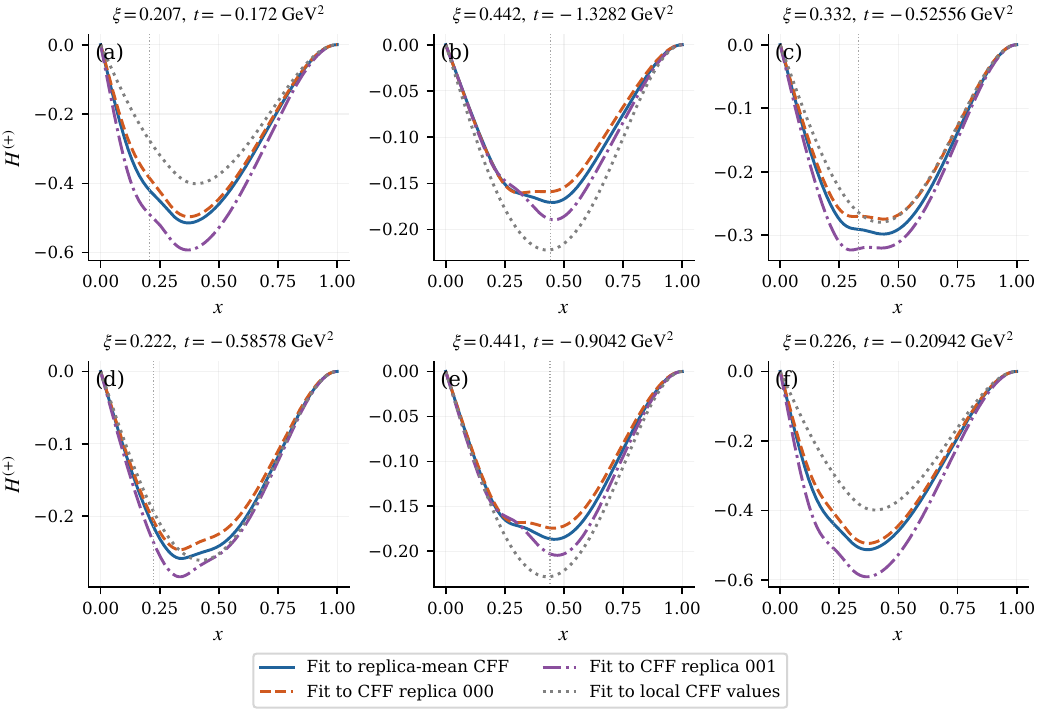}
 \caption{GPDs reconstructed from experimental CFF inputs at
 $\Qz=2~\mathrm{GeV}^2$. Blue curves are GPDs fitted to the
 pointwise mean of 32 CFF replica functions; orange and purple
 curves are GPDs fitted to CFF replicas 1 and 2.
 These three fits use 20 additional optimization iterations
 from the same starting network.
 Gray dotted curves show the separate local-CFF fit, which
 uses the 160 unperturbed local CFF values directly, without
 intermediate CFF smoothing. All six preselected $(\xi,t)$ points are shown;
 vertical dotted lines mark $x=\xi$. The two replica fits
 illustrate the input sensitivity.}
 \label{fig:family}
\end{figure*}

\begin{figure*}[t]
 \includegraphics[width=\textwidth]{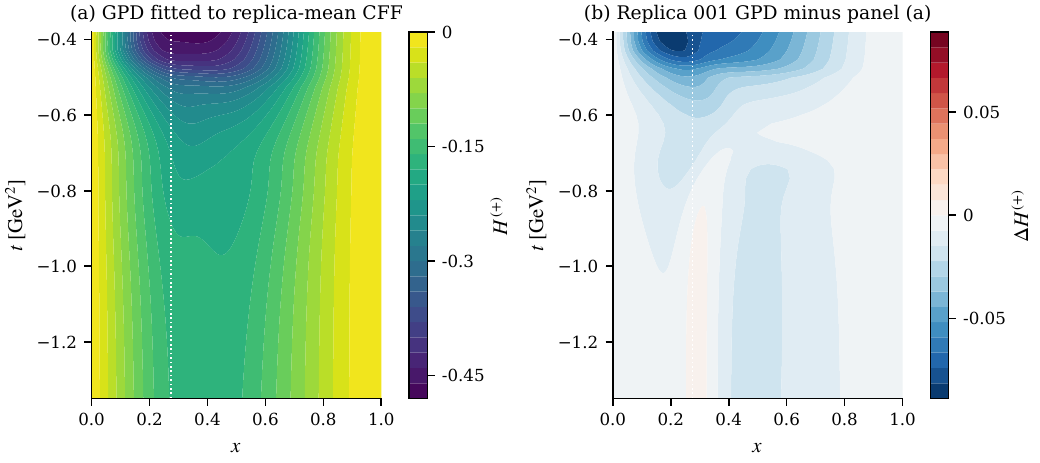}
 \caption{GPD fields at $\Qz=2~\mathrm{GeV}^2$ and
 $\xi=0.275$, after 20 additional optimization iterations.
 (a) The GPD fitted to the pointwise mean of the 32 CFF
 replicas. (b) The GPD fitted to CFF replica 2 minus
 the GPD shown in panel (a). The panels use different color scales, and
 the white dotted line marks the crossover.}
 \label{fig:experiment_surface}
\end{figure*}

Figure~\ref{fig:family} displays the replica-mean CFF fit and
the GPDs fitted to CFF replicas 1 and 2 after 20 additional
iterations, at six preselected $(\xi,t)$ points.
The continuous neural fields allow the same comparison away
from those slices; Fig.~\ref{fig:experiment_surface} shows an
$(x,t)$ plane and the signed difference between the GPD
fitted to replica 2 and the GPD fitted to the mean CFF.


The displayed fields can take negative values in the charge-weighted
convention at finite $t$; no forward PDF boundary or positivity
condition is imposed in these fits.

\subsection{Dependence on optimization iterations and CFF residuals}
For each CFF replica $r$, compare its reconstructed GPD with
the replica-mean CFF fit after the same number $b$ of
additional outer iterations:
\begin{equation}
 D_r^{(b)}(x,\xi,t)
 =\widehat{\Hp}_r^{(b)}(x,\xi,t)
  -\widehat{\Hp}_{\mathrm{mean}}^{(b)}(x,\xi,t).
 \label{eq:response}
\end{equation}
Here $\widehat{\Hp}_{\mathrm{mean}}^{(b)}$ denotes the
replica-mean CFF fit. The RMS of this difference on the six
displayed full-$x$ slice grids increases from 0.0091 to
0.0131 for replica 1 and from 0.0070 to 0.0337 for
replica 2 as $b$ increases from 5 to 20.
These RMS differences are in GPD units, evaluated on the displayed
slice grids.

\begin{figure*}[t]
 \includegraphics[width=\textwidth]{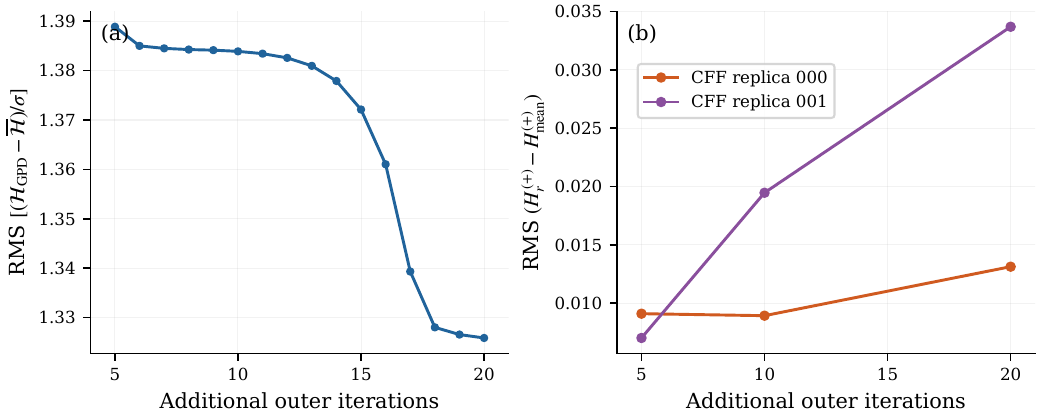}
 \caption{Dependence of the experimental reconstruction on
 the number of additional outer optimization iterations.
 (a) RMS difference between the $\ReH$ predicted by the
 replica-mean CFF fit and the mean CFF target, divided
 pointwise by the original quoted errors.
 (b) RMS difference between each replica's GPD and the
 replica-mean CFF fit, evaluated after the same number of
 iterations on the six full-$x$ slice grids.}
 \label{fig:budget}
\end{figure*}

Over the same iterations, the RMS residual between the
replica-mean CFF fit's predicted $\ReH$ and its target CFF
mean decreases from 1.39 to 1.33, with each residual
divided by the original quoted error.
After 20 iterations, its corresponding RMS residual against
the 160 unperturbed local CFF values is 2.17.
Figure~\ref{fig:budget} shows that a modest change in CFF fit
quality accompanies a substantial change in the field response,
with no plateau reached by 20 iterations.

For each replica, we also compare its predicted real-CFF
change relative to the replica-mean CFF fit with its input
real-CFF change relative to the mean CFF target. Taking
the ratio of their norms, with both changes divided by
the original quoted errors, gives 0.34 and 0.43 for
replicas 1 and 2 after 20 additional iterations.
The output and input perturbations differ in direction as well as
in norm.

The residual relative to the mean CFF can be separated from the
dispersion among the 32 CFF replicas.
Let $\bm R_r$ be the 32 CFF vectors in the source sign
convention, $\overline{\bm R}$ their mean,
$W_\sigma=\operatorname{diag}(1/\sigma_i)$, and $\bm g$
a fixed GPD prediction. Then
\begin{align}
 \frac{1}{mn}\sum_{r=1}^{m}
  \|W_\sigma(\bm g-\bm R_r)\|^2
 &=\frac1n\|W_\sigma(\bm g-\overline{\bm R})\|^2\nonumber\\
 &\quad+\frac{1}{mn}\sum_{r=1}^{m}
 \|W_\sigma(\bm R_r-\overline{\bm R})\|^2 .
 \label{eq:decomposition}
\end{align}
For the replica-mean CFF fit after 20 additional iterations,
with $m=32$ replicas and $n=160$ coordinates, the two terms are
$1.758+0.304=2.062$.
The residual relative to the mean CFF dominates the dispersion
among replicas.

Independent finite-moment checks at all 79 measured $t$
values give combined normalized residuals of 0.29\%,
0.28\%, and 0.28\% for the replica-mean CFF fit and the
fits to replicas 1 and 2, respectively, after 20 additional
iterations.
\section{Discussion and outlook}\label{sec:discussion}
The exact-input tests support a concrete methodological result:
the reference-scale neural field, coupled to NLO evolution and
matching, can recover a nontrivial charge-weighted GPD without
being supplied its pointwise values or true Mellin coefficients.
The disjoint CFF check also tests prediction across a multiscale
kinematic chart. The conditional FiLM representation,
free-coefficient moment objective, and constrained optimization
together define a reconstruction procedure designed to extract
information from experimentally obtained CFFs, distinct from previous
neural approaches and from the broader non-neural GUMP analysis
\cite{Guo:2025gpd}.

The reconstruction has NLO perturbative accuracy in the specific
sense defined in Sec.~\ref{sec:qcd_accuracy}: the learned initial
field enters a leading-twist CFF map with two-loop evolution,
one-loop hard coefficients, and two-loop coupling evolution in a
common scheme. This accuracy is a property of the factorized
calculation used in every fit and is distinct from the closure error,
which measures how well a specified input field is recovered.

The noiseless and noisy cases answer different questions.
Exact-input closure measures the ability of a specified inverse
procedure to reconstruct a known function under controlled
conditions. The noisy examples show how that procedure responds
when uncertain integral observations are supplied.
The experimental family then demonstrates the same propagation
at the level of complete CFF functions.
Together these results separate accurate CFF reproduction from
determination of the GPD itself.

Several extensions follow directly. Additional generating truths and
initializations would broaden the closure study; a denser ideal
lattice and higher moment spins would tighten the constraints; and
measured imaginary CFFs, a full experimental covariance, and
simultaneous extraction of the other leading-twist GPDs would extend
the experimental application. Moments of $H$ alone do not determine
the quark angular momentum, which requires $H$ and $E$ together
\cite{ji-1997-gauge-invariant-decomp}. A larger set of propagated
replicas under one fixed reconstruction rule would characterize that
rule's Monte Carlo distribution, reported together with its dependence
on iteration count and initialization.

Additional physical information can be included as an
independent residual block. Forward parton distributions,
lattice moments, or complementary observables may constrain
directions weakly seen by the present CFFs. The
forward-boundary comparison in Appendix~\ref{app:pdf}
illustrates this: improvement of one functional need not
improve the entire GPD.

Symbolic analysis of fitted moments offers another descriptive
extension. A restricted polynomial-support hypothesis can be
proposed and then refitted with its remaining coefficients free
through the same physics map. No such restriction is used in
the results reported here.

\section{Conclusion}
We have constructed and applied a differentiable NLO neural
GPD reconstruction method with coupled quark and gluon fields,
reference-scale parameterization, exact discrete symmetries,
and free-coefficient finite-spin polynomiality constraints.
In noiseless tests it recovers the charge-weighted
quark field at percent-level accuracy for the specified
generating function, with disjoint multiscale CFF prediction.
At experimental coordinates, introducing a fixed perturbation
at the reported error scale produces much larger and
initialization-dependent field errors.

For experimentally extracted real CFFs, the same framework
produces continuous GPD surfaces whose response to input
fluctuations, residual mismatch to the mean CFF, and dependence on
the number of optimization iterations are all quantified.

\section*{Data and software availability}
The software and saved analysis products are archived as
version 0.3.0 on Zenodo~\cite{Keller2026Software}.
The source repository is available at
\url{https://github.com/zetanaut/GPD-inverse-integral}.
The manuscript source package includes the numerical arrays
used for all figures, plotting scripts, and a source-hash
provenance manifest. Figure generation reads saved results.
Reproducing the full inverse trajectories additionally
requires the corresponding operator inputs and saved run
configurations identified in that manifest.

\appendix
\section{Analytic closure generator}\label{app:truth}
Table~\ref{tab:truth} specifies every nonzero coefficient
in Eq.~\eqref{eq:truth}.
All omitted coefficients and additional $D$-term amplitudes
are zero. The slopes $b$ have units $\mathrm{GeV}^{-2}$.
The amplitudes are dimensionless.

\begin{table}[h]
 \caption{Nonzero coefficients $(a,b)$ of the closure generator.
 A dash denotes an absent term.}\label{tab:truth}
 \begin{ruledtabular}
 \begin{tabular}{ccccc}
 $i$ & $k$ & $\Sigma:\ (a,b)$ & $T:\ (a,b)$ & $H_g:\ (a,b)$\\ \hline
 0&0&$(6,0.65)$&$(2.4,0.85)$&$(1.5,0.55)$\\
 0&1&$(0.90,0.90)$&$(-0.35,0.60)$&$(0.22,0.75)$\\
 1&0&$(-0.30,1.05)$&$(0.12,1.10)$&$(-0.045,0.95)$\\
 1&1&$(0.20,0.70)$&$(-0.08,0.90)$&---\\
 1&2&$(0.05,0.80)$&---&$(0.022,0.60)$\\
 2&0&$(0.08,1.20)$&---&$(0.02,1.10)$\\
 2&1&---&$(0.025,1.30)$&---
 \end{tabular}
 \end{ruledtabular}
\end{table}

\section{Forward-boundary information test}\label{app:pdf}
A separate comparison adds 36 synthetic forward
constraints: three reference-scale channels at 12 logarithmic
$x$ nodes over $0.02\leq x\leq0.8$.
Their declared full covariance is
\begin{align}
 C_{ic,jd}&=s_{ic}s_{jd}
  e^{-|\ln x_i-\ln x_j|/0.7}\,R_{cd},\nonumber\\
 s_{ic}&=0.10\max\left(|f_c(x_i)|,\,
                  0.05\max_k|f_c(x_k)|\right),
\end{align}
where $f_c$ is the generating forward field.
The channel correlation matrix has unit diagonal,
$R_{\Sigma T}=R_{T\Sigma}=0.5$, and zero quark--gluon
off-diagonal entries.
The extra loss is the whitened squared residual, normalized
by the same 160-row count as the CFF term, with unit
coefficient in the joint fit.

Two scalar diagnostics, evaluated at $\xi=0.275$,
$t=-0.363~\mathrm{GeV}^2$, and $\Qz=2~\mathrm{GeV}^2$, are
\begin{align}
 A&=5\int_{0.2}^{0.4}\Hp(x,\xi,t)\,\dd x,\nonumber\\
 B&=\int_0^1x\,\Hp(x,\xi,t)\,\dd x.
\end{align}
Here $A$ is a bin average and $B$ a positive-$x$ moment.
They are evaluated after the fit and are not part of the objective.
Table~\ref{tab:pdf} reports absolute relative errors for
these targets and relative field errors on the observed
and wider charts.

\begin{table}[t]
 \caption{Comparison of CFF-only and joint CFF--forward-boundary fits.
 All entries are errors in percent; target columns use absolute
 relative error. Half noise denotes inputs perturbed
 at half the quoted CFF error scale. This case is separate
 from the full-noise example in Table~\ref{tab:noisy}.}\label{tab:pdf}
 \setlength{\tabcolsep}{3pt}
 \begin{ruledtabular}
 \begin{tabular}{llrrrr}
 Input & Fit & $H_{\mathrm{obs}}^{(+)}$ & $A$ & $B$ & $H_{\mathrm{chart}}^{(+)}$\\ \hline
 Exact & CFF & 0.92 & 0.49 & 0.87 & 0.75\\
 Exact & CFF+PDF & 0.47 & 0.68 & 0.25 & 0.54\\
 Half noise & CFF & 11.5 & 4.99 & 1.28 & 28.6\\
 Half noise & CFF+PDF & 10.1 & 3.54 & 3.24 & 38.1
 \end{tabular}
 \end{ruledtabular}
\end{table}

With exact CFF inputs, the joint fit improves the wider-chart
field error while keeping both target errors below 1\%.
For the half-noise case it improves the observed-chart field
error and target $A$, but worsens target $B$ and the wider
field. Improvement in the measured region or in one scalar
diagnostic can thus coexist with deterioration elsewhere.

\clearpage
\bibliographystyle{apsrev4-2}
\bibliography{references}
\end{document}